\documentclass{elsart}
\usepackage{multicol}
\usepackage{graphicx}

\makeatletter
\renewcommand{\@journal}{Optics Communications}
\makeatother

\newcommand{\figwidth}{0.4\textwidth}

\begin{document}

\begin{frontmatter}

\title{Spatial coherence of thermal near fields}

\author{C. Henkel\thanksref{1}}
\thanks[1]{Corresponding author. Telephone: (49)331.977 14 98,
facsimile: (49)331.977 17 67,
electronic mail:
\makeatletter Carsten.Henkel@quantum.physik.uni-potsdam.de\makeatother
}

\address{Institut f\"ur Physik, Universit\"at Potsdam, 14469 Potsdam,
Germany}

\author{K. Joulain, R. Carminati, J.-J. Greffet}
\address{Laboratoire d'Energ\'etique Mol\'eculaire et Macroscopique,
Combustion\thanksref{2},
Ecole Centrale Paris, 92295 Ch\^atenay-Malabry cedex, France}
\thanks[2]{Unit\'e propre de recherche no. 288 du Centre National
de la Recherche Scientifique}

\date{20 july 2000}

\begin{abstract}
We analyze the spatial coherence of the electromagnetic
field emitted by a
half-space at temperature $T$ close to the interface.  An asymptotic
analysis allows to identify three different contributions to the
cross-spectral density tensor in the near-field regime.  It is shown
that the coherence length can be either much larger or much shorter
than the wavelength depending on the dominant contribution.
\\[3mm]
\end{abstract}

\end{frontmatter}

\begin{multicols}{2}

\section*{Introduction}

The typical textbook example of an incoherent source is a thermal
source. From the point of view of temporal coherence, its spectrum
is given by Planck's function and modified by its emissivity. For
usual sources, the emissivity is a smooth function of frequency.
Thus,
the spectral width is usually on the order of the peak frequency of
Planck's function. From
the point of view of spatial coherence, a thermal source is often
assumed to be delta correlated. Yet, an exact form of the
cross-spectral density tensor has been derived for a blackbody
radiator
and it has been shown that the spatial coherence length is
$\lambda/2$ \cite{MandelWolf}.
These exact results seem to support the statement that a
thermal source is incoherent.  Yet, one has to analyze more
carefully the problem when dealing with a real thermal source. In this
paper, we consider a source that consists of a half space filled with
a lossy material at temperature $T$. We are interested in the emitted
field so that we assume that there are no other sources. Thus there is
no incident radiation illuminating the sample. Note in particular that
this is not an equilibrium situation.

Since we explicitly introduce a model for the source, the emitted
field contains evanescent waves in the vicinity of the
interface. These evanescent waves are not taken into account when
dealing with blackbody radiation. Yet, they modify the coherence
properties of the source in the near field as was shown in
\cite{Greffet99}.
The effect is particularly striking if a resonant surface wave is
excited. It has been shown that the coherence length can be either
much larger than the wavelength or much shorter than $\lambda/2$
close to the
surface. Temporal coherence is also dramatically modified. For
example, the emitted radiation is almost monochromatic when a
surface wave is excited \cite{Shchegrov00}. These results were
obtained using a direct
calculation of the field emitted by a half-space in the framework of
fluctuation electrodynamics
\cite{Lifshitz56,Polder71,Schwinger78,Rytov3}.

The subject of this paper is to analyze the spatial coherence of the
emitted field by means of an asymptotic evaluation of the
cross-spectral density tensor in the near-field limit
(interface-detector distance small compared to the wavelength). This
analysis permits to retrieve the properties reported in
\cite{Greffet99} and
yields insight into the physical mechanism responsible for
these effects. We are thus able to identify all the
possible contributions to the cross-spectral density tensor :
thermally excited surface plasmons, skin-layer currents and
small-scale polarization fluctuations. We show that to a good
approximation, the sum of these three asymptotic contributions
coincides with the exact result. We obtain different characteristic
behaviours that vary in accordance with the dominant term. Surface
waves
such as surface plasmon-polaritons or surface phonon-polaritons yield
long-range spatial coherence on a scale of the surface wave
propagation
length which may
be much larger than the wavelength when aborption is small. On the
contrary, skin-layer currents and small-scale polarization
fluctuations lead to a much shorter spatial coherence length that only
depends on the distance to the interface. A surprising consequence of
this property is that the macroscopic theory of radiometry may be
extended into the mesoscopic regime insofar as emission is concerned.
Note however that this
conclusion is based on the assumption of a local medium.
The ultimately limiting scale is thus given by the electron
screening length or the electron Fermi wavelength, whatever is larger
\cite{Feibelman82,Ford84}.

\section{Overview}

\subsection{Radiation emitted by a thermal source}

In this section, we review the source theory approach we use for the
computation of the thermal electromagnetic field
\cite{Lifshitz56,Polder71,Schwinger78,Rytov3}. We focus on the
radiation in the vacuum close to a source that we model as a linear
dielectric with dielectric function $\varepsilon({\bf r}; \omega)$.
The frequency dependence will not be indicated explicitly in the
following since we calculate quantities at fixed frequency (or,
equivalently, at fixed wavelength $\lambda =
2\pi c / \omega$. The source radiates because it contains a
fluctuating polarization field ${\bf P}( {\bf r} )$. The
spectral density of this field is characterized by the
cross-correlation
tensor $S_P^{ij}( {\bf r}_1, {\bf r}_2 )$ that,
according to the fluctuation-dissipation theorem
\cite{Lifshitz56,Polder71,Agarwal75a,Scheel98a}, is given by
\begin{equation}
S_P^{ij}( {\bf r}_1, {\bf r}_2 )
=
\frac{ 2\hbar \varepsilon_0 {\rm Im}\, \varepsilon({\bf r}_1)
}{
{\rm e}^{ \hbar \omega / k_B T } - 1
}
\delta^{ij}
\delta( {\bf r}_1 - {\bf r}_2 )
\label{eq:FD-polarization}
\end{equation}
The Kronecker $\delta^{ij}$ and the spatial $\delta$-function in this
formula
follow from the assumption that the dielectric function is isotropic
and
local. We have taken the normal-ordered form for the polarization
field
spectrum since we are ultimately interested in the electromagnetic
field
measured by a photodetector (given by normally-ordered field operators
\cite{MandelWolf,Agarwal75a}). The electric field ${\bf E}( {\bf r} )$
radiated by the polarization
${\bf P}( {\bf r} )$ is now given by the Green function for the
source geometry
\begin{equation}
E_i( {\bf r} ) =
\int\limits_V
\!{\rm d}{\bf r}' \sum_j\,
G^{ij}( {\bf r}, {\bf r}' )
P_j( {\bf r}' )
\label{eq:E-and-Green}
\end{equation}
where $V$ is the volume of the source, \emph{i.e.}, the domain where
${\rm Im}\,\varepsilon( {\bf r}' )$ is nonzero according
to~(\ref{eq:FD-polarization}).
All quantities in~(\ref{eq:E-and-Green}) are understood as
temporal Fourier transforms at frequency
$\omega$. The coherence function $W^{ij}( {\bf r}_1, {\bf r}_2 )$
of the electromagnetic field is now obtained as a
thermal average of~(\ref{eq:E-and-Green}), using the polarization
spectrum~(\ref{eq:FD-polarization}). One obtains
\cite{Greffet99,Shchegrov00}
\begin{eqnarray}
W^{ij}( {\bf r}_1, {\bf r}_2 ) & = &
\frac{ 2 \hbar \varepsilon_0 }{
{\rm e}^{ \hbar \omega / k_B T } - 1
}
\sum_k
\int_V\!{\rm d}{\bf r}'
{\rm Im}\,\varepsilon( {\bf r}' ) \times {}
\nonumber\\
&& {} \times
G^{ik*}( {\bf r}_1, {\bf r}' )
G^{jk}( {\bf r}_2, {\bf r}' )
\label{eq:SE-and-Green}
\end{eqnarray}
The problem is now to evaluate this expression analytically and to
obtain
an estimate for its dependence on the separation ${\bf s} \equiv {\bf r}_2
- {\bf r}_1$ between the observation points.

\vspace*{5mm}
\centerline{
\resizebox{\figwidth}{!}{
\includegraphics*{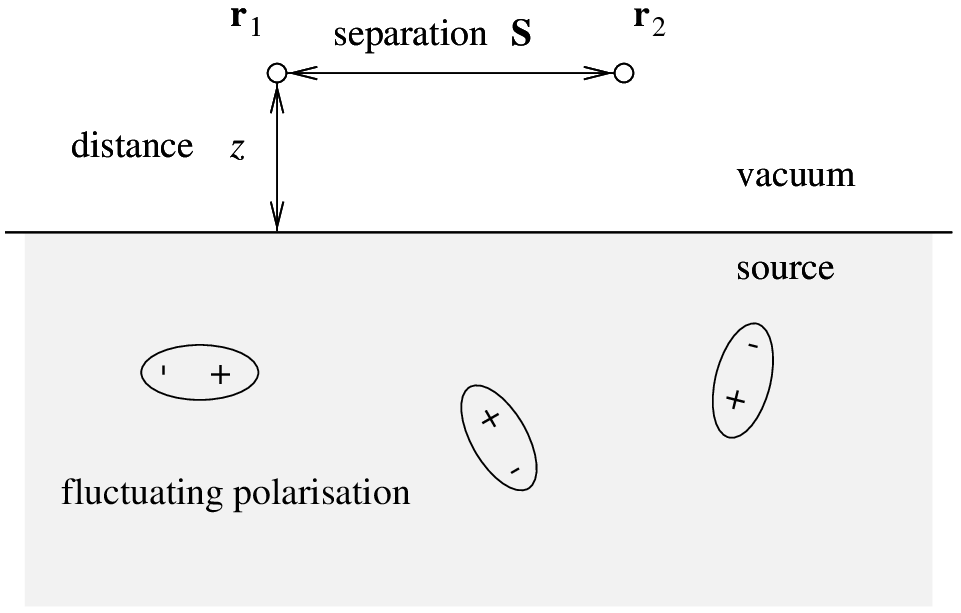}}}
\begin{quote}\small
{Fig.1:}
Model geometry for a planar source.
\refstepcounter{figure}
\end{quote}

To proceed in the calculation, we focus on the simplified geometry
shown in figure~1:
an infinite half-space with uniform dielectric constant
$\varepsilon$, separated by the plane $z=0$ from the empty half-space
$z > 0$.
For this arrangement, the Green tensor is explicitly known as a
spatial Fourier
transform with respect to the lateral separation ${\bf S} = (s_x,
s_y) \equiv
{\bf R}_2 - {\bf R}_1$. Details may be
found in \cite{Agarwal75a,Maradudin75,Henkel98a} and in
appendix~\ref{a:Green}.
As to be expected for this source geometry, the electric coherence
tensor
depends on the distances $z_1, z_2$ of the observers and their lateral
separation ${\bf S}$. For simplicity, we put in the following
$z_1 = z_2 = z$. We also
normalize the coherence tensor $W^{ij}$ to its value for ${\bf r}_1
= {\bf r}_2$ in the case of blackbody radiation
\begin{equation}
W_{bb} = \frac{ 2 \hbar k^3 }{ 3 \pi \varepsilon_0
( {\rm e}^{ \hbar \omega / k_B T } - 1 ) }
\label{eq:SE-bb}
\end{equation}
where as usual $k = \omega/c$.
As outlined in appendix~\ref{a:Green},
we thus get the following expression for the spatial Fourier transform
of the coherence tensor
\begin{eqnarray}
w^{ij}( {\bf S}, z ) & = & \frac{ W^{ij}( {\bf S}, z ) }{ W_{bb} }
\nonumber
\\
& = &
\int\!\frac{ {\rm d}^2{\bf K} }{ (2\pi)^2 }
{\rm e}^{ {\rm i} {\bf K} \cdot {\bf S}
- 2 z {\rm Im}\,\gamma }
w^{ij}( {\bf K} )
\label{eq:def-w-ij}
\end{eqnarray}
where ${\bf K}$ denotes a wave vector parallel to the interface and
$( {\bf K}, \gamma)$ is the vacuum wave vector of a plane wave emitted
by the source. The tensor $w^{ij}( {\bf K} )$ is given in
appendix~\ref{a:Green}, eq.(\ref{eq:formula-w-ij}).
The integration over ${\bf K}$
in~(\ref{eq:def-w-ij}) also includes wave vectors $|{\bf K}| > k$,
describing evanescent waves the source excites in the vicinity
of the interface (the quantity $\gamma$ is then purely imaginary with
positive imaginary part).

\subsection{Near field coherence function}

In this subsection, the typical behaviour of the field coherence
function
is discussed. We identify several distance regimes showing a very
different behaviour of the lateral coherence function. Analytical
approximations for the coherence function are deferred to the next
section.

\begin{figure*}
\centerline{
\resizebox{0.4\textwidth}{!}{
\includegraphics*{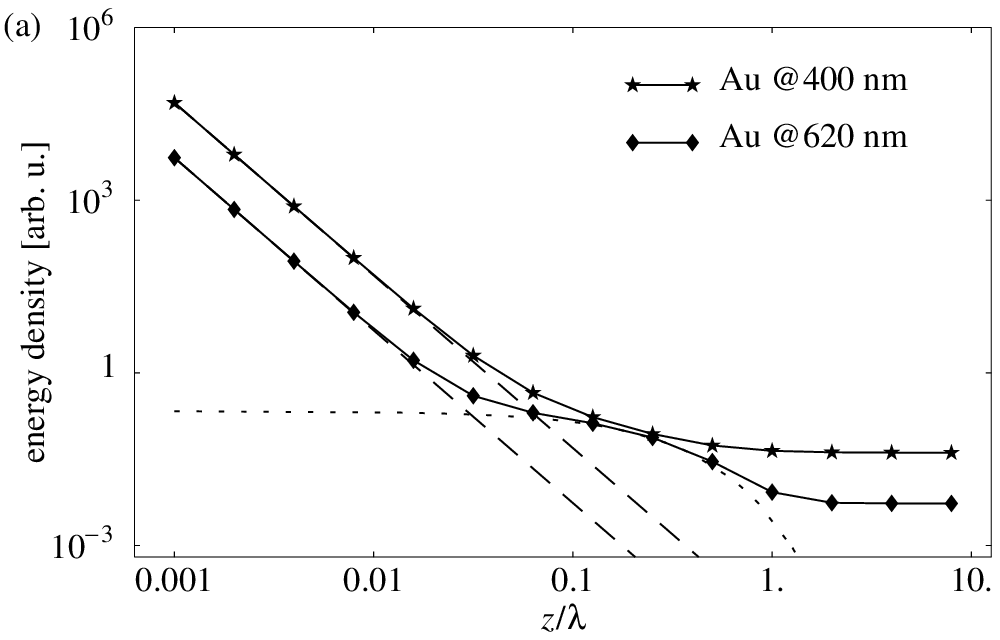}}
\hspace*{1mm}
\resizebox{0.4\textwidth}{!}{
\includegraphics*{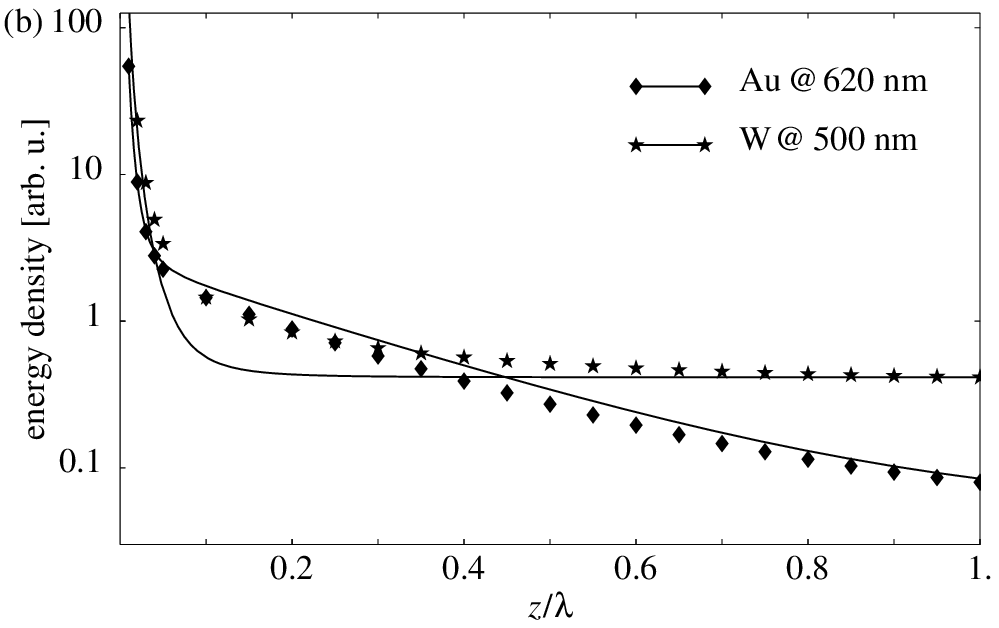}}}
\caption[]{\small
Energy density ${\rm Tr}\,w^{ij}( {\bf S} = {\bf 0}, z )$
vs.\ distance from a metal surface.
Dots: numerical evaluation of the integral~(\ref{eq:def-w-ij}),
solid lines: sum of the asymptotic approximations discussed
in the text. In the far field, the numerically computed value
is taken.
\\
(a): log-log scale for gold at $\lambda = 620\,$nm
($\varepsilon = -9.3 + 1.2\,{\rm i}$)
and at $\lambda = 400\,$nm
($\varepsilon = -1.1 + 6.5\,{\rm i}$).
The dielectric constants are extracted from \cite{Palik}.
Dashed line: $1/z^3$ power law dominating the extreme near field;
dotted line: exponentially decaying contribution of excited
surface modes.
\\
(b): log-linear scale, showing
the exponentially decaying surface plasmon contribution
for gold at $\lambda = 620\,$nm.
For comparison, the case of tungsten at $\lambda = 500\,$nm
is shown where no plasmon resonance is found
($\varepsilon = 4.4 + 18\,{\rm i}$).
}
\end{figure*}

In figure~2 is shown the `energy density'
(the trace of the coherence tensor at coinciding positions) above
a metal surface in double logarithmic scale.
One observes a strong increase with respect to the far field
energy density when the distance $z$ is smaller than the wavelength.
For moderate distances $z \le \lambda$, the energy density is
dominated by an exponentially increasing contribution [cf.\
fig.2(b)]. This is due to the excitation of surface
plasmon resonances, whose contribution is calculated analytically
in subsection~\ref{s:plasmon}. The other curve in fig.2(b) shows
the energy density for the case of tungsten with
${\rm Re}\,\varepsilon > -1$ where
no surface mode exists and no exponential increase is found.
For small distances $z \ll \lambda$,
the energy density follows a $1/z^3$ power law (`static limit') that
is
discussed in subsection~\ref{s:xnf}. The prefactor of this power law
involves the imaginary part of the electrostatic reflection coefficient
Im[$(\varepsilon - 1)
/ (\varepsilon + 1)$]. The second curve in fig.2(a) illustrates the
resonantly
enhanced energy density for a wavelength where
${\rm Re}\,\varepsilon \approx -1$. The `static limit' contribution
then overwhelms that of the plasmon resonance.

In figure~3, we show the normalized lateral
coherence function at chosen distances from the interface.
In the far field [plot 3(a)], the coherence length is $\lambda/2$,
and the coherence function the same as for the blackbody field
($(\sin ks ) / ks $ behaviour). This is not
surprising since at large distances $z \gg \lambda$, only propagating
plane waves radiated into
the vacuum half space contribute to the field.

When the surface plasmon excitation dominates the field energy
($z \le \lambda$), the
field coherence extends over much longer distances [plot 3(b)]. This
is because of the weak damping of the plasmon modes in this case. We
show below (subsection~\ref{s:plasmon}) that the coherence length
is indeed given by the plasmon
propagation length. The figure also shows that the field is strongly
polarized perpendicular to the interface, as is the surface plasmon
mode.

At distances $z \ll \lambda$ even closer to the source,
the field coherence length gets shorter again [plot 3(c)]. We show
below
that in this regime, the field behaves as if it was quasi-static
(subsection~\ref{s:xnf}).
This leads to a lateral coherence length equal to the vertical
distance
$z$ from the interface and hence much shorter than the wave length.
We thus find the surprising result that thermal near fields
have no lower limit in their coherence length, as long as the
dielectric
function of the source may be taken as local.

\begin{figure*}
\centerline{
\resizebox{0.38\textwidth}{!}{
\includegraphics*{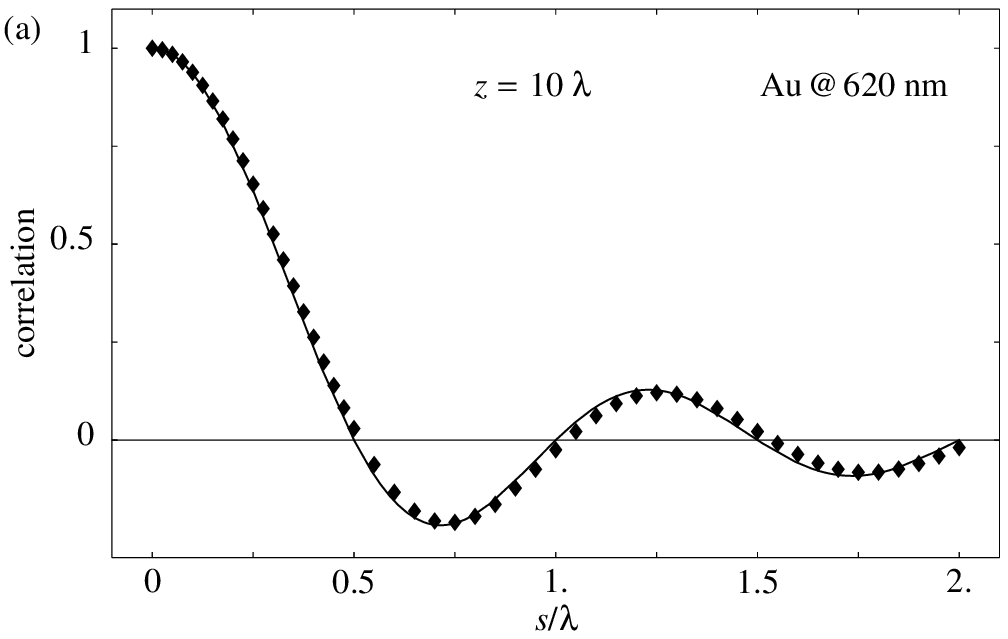}}
\hspace*{1mm}
\resizebox{0.4\textwidth}{!}{
\includegraphics*{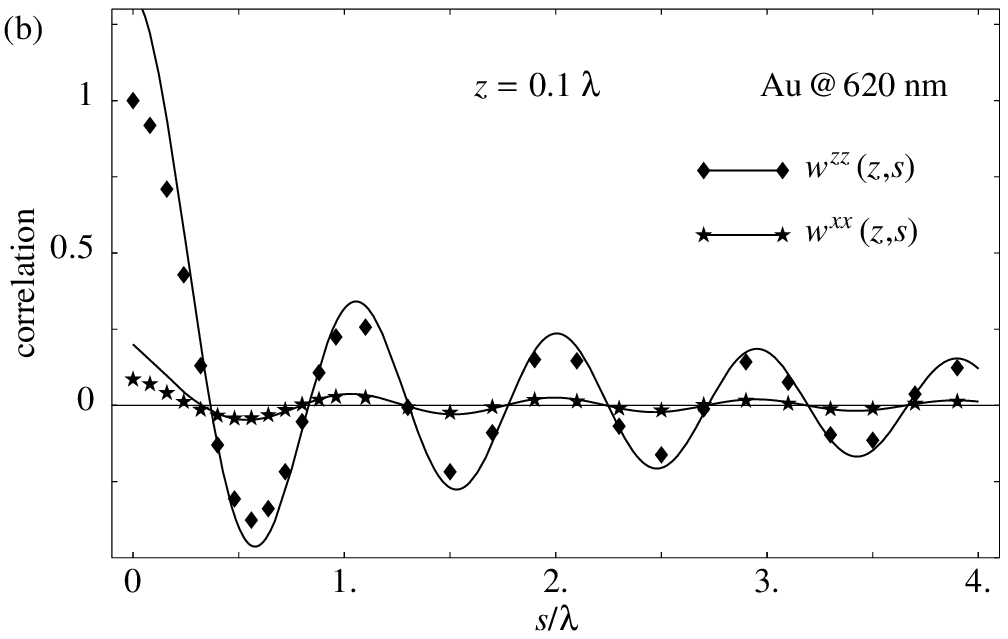}}}
\centerline{
\resizebox{0.45\textwidth}{!}{
\includegraphics*{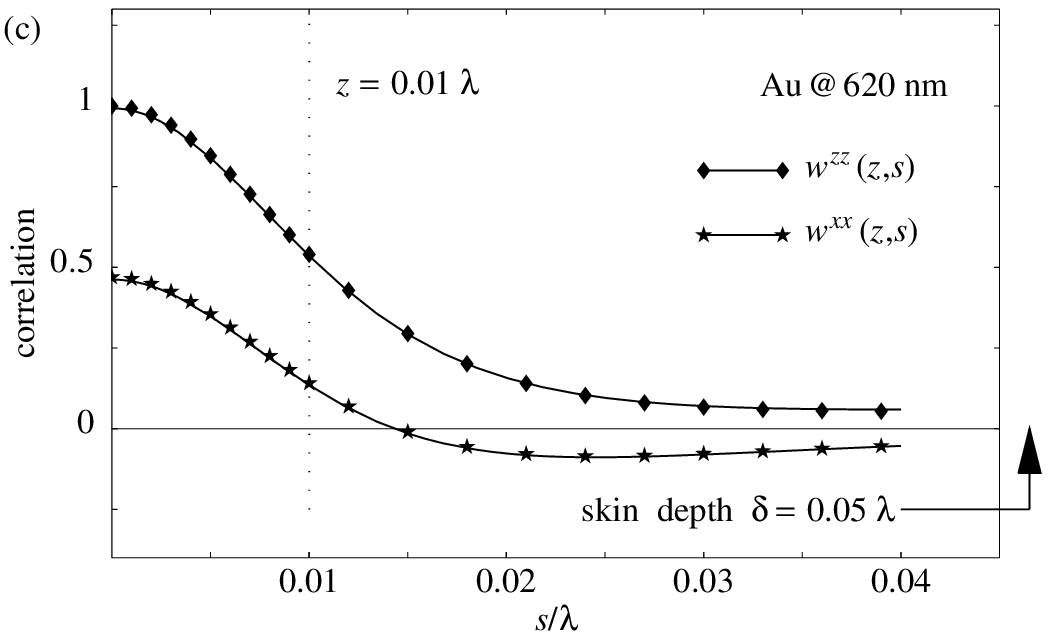}}}
\caption[]{\small
Normalized lateral coherence functions
for three fixed distances $z$, plotted vs.\ the lateral separation
$s = |{\bf S}|$. All plots are for a gold surface at $\lambda =
620\,$nm.
Dots: numerical evaluation of~(\ref{eq:def-w-ij}),
solid lines: analytical approximations discussed in the text.
The numerically computed values were used to normalize all curves.
\\
(a): far field regime $z = 10\,\lambda$. The trace of the
coherence tensor is plotted, normalized to its value for $s=0$. Solid
line:
free space coherence function $\sin(ks)/(ks)$.
\\
(b): plasmon dominated regime $z = 0.1\,\lambda$. The components
$w^{xx}$ and $w^{zz}$ are plotted, normalized to (the numerically
computed) $w^{zz}( {\bf S} = {\bf 0}, z )$.
\\
(c) static regime $z = 0.01\,\lambda$. The components
$w^{xx}$ and $w^{zz}$ are plotted and normalized as in plot~3(b).
The solid curve only contains the extreme near-field
contribution~(\ref{eq:xnf-contribution}).
}
\end{figure*}

One might finally ask whether the skin depth $\delta$ [defined
in~(\ref{eq:def-skin-depth})] is relevant for
the radiation emitted by a metallic source. This question is
discussed in
subsection~\ref{s:skin} where we show that in the regime
$\delta \ll z \ll \lambda$, a different power law ($\propto 1/z^2$)
governs the energy density (see fig.4(a) below). The lateral
coherence
behaves similar to the static regime $z \ll \delta$, however, as
shown in
fig.4(b) below.

\section{Analytical approximations}

\subsection{Plasmon contribution}
\label{s:plasmon}

It is well known that a dielectric-vacuum interface supports surface
plasmon polariton (or phonon polariton) modes provided the dielectric
constant satisfies
${\rm Re}\,\varepsilon < -1$ \cite{AgranovichMills,Raether}.
These surface modes propagate
parallel to the surface with a wave vector $K_{\rm pl}$ and are
exponentially localized
in the direction perpendicular to the interface. In addition, if
there are
losses in the dielectric, the propagation parallel to the interface
is damped which may be described by a
nonzero imaginary part of $K_{\rm pl}$. Mathematically, we obtain the
plasmon
dispersion relation by identifying the poles of the transmission
coefficients
$t_\mu$ ($\mu = {\rm s,\,p}$) as a function of the wave vector $K$.
Only the p-polarization (magnetic field perpendicular to the plane of
incidence) gives a pole at the (complex) position
\begin{equation}
K_{\rm pl} = k \sqrt{ \frac{ - \varepsilon }{ - \varepsilon - 1 } },
\quad
{\rm Re}\,K_{\rm pl}, \:  {\rm Im}\,K_{\rm pl} > 0
\label{eq:K-plasmon}
\end{equation}
The plasmon pole shows up as a sharp peak when
the integrand $w^{ij}( {\bf K} )$ in~(\ref{eq:def-w-ij})
is plotted as a function of the lateral wave vector magnitude $K$
[cf.\ eq.(\ref{eq:formula-w-ij})]. This
suggests that we get a good approximation to the plasmon contribution
by taking slowly varying terms outside the integral and evaluating the
pole contribution by contour integration. For example, the
denominator of the
$|t_{\rm p}|^2$ transmission coefficient may be approximated as
\begin{equation}
\frac{ 1 }{ | \varepsilon \gamma + \gamma_2 |^2 }
\approx
\frac{ 4 |\varepsilon|^2 }{ |\varepsilon + 1| \, |\varepsilon - 1|^2 }
{\rm Im}\frac{ 1 }{ K^2 - K_{\rm pl}^2 }
\label{eq:trick-tp}
\end{equation}
where $\gamma = \gamma(K)$ and $\gamma_2$ are the vertical wave vector
components above and below the interface.
It is essential for the contour integration to work that one
expresses the
absolute square on the left hand side as the imaginary part of an
analytic
function of $K$ (right hand side).

It is easily checked from~(\ref{eq:formula-w-ij}) that the trace of
$w^{ij}( {\bf K} )$ only depends on the magnitude $K$ of the lateral
wave
vector ${\bf K}$. The integration over the angle between ${\bf K}$ and
${\bf S}$ therefore gives
\begin{eqnarray}
&&
{\rm Tr} \, w^{ij}( {\bf S}, z ) =
\nonumber\\
&&
\int_0^\infty \!\frac{ K\,{\rm d}K }{ 2 \pi }
J_0( K s )
\,{\rm e}^{ - 2 z {\rm Im}\,\gamma}
{\rm Tr} \, w^{ij}( {\bf K} )
\end{eqnarray}
where $J_0( \cdot )$ is the ordinary Bessel function and $s = |{\bf
S}|$.
The individual tensor components also involve Bessel functions
$J_2( K s )$, as discussed in appendix~\ref{a:plasmon-polarization}.
The integration over $K$ may be done using the
identity~(\ref{eq:J0-theorem}) proven in appendix~\ref{a:contour}.
The diagonal elements of the coherence tensor finally take the
suggestive form
\begin{eqnarray}
w^{ii}( {\bf S}, z ) &\approx&
C_{\rm pl}
\,{\rm e}^{ - 2 \kappa_{\rm pl} z }
g^{i}( K_{\rm pl} s )
\label{eq:plasmon-contribution}
\\
g^{\Vert}( u ) & = &
\frac12
{\rm Re}\!\left[
H_0( u )
- H_2( u )
- \frac{ 4 {\rm i} }{ \pi u^2 }
\right]
\label{eq:plasmon-contribution-x}
\\
g^{\perp}( u ) & = &
\frac12
{\rm Re}\!\left[
H_0( u )
+ H_2( u )
+ \frac{ 4 {\rm i} }{ \pi u^2 }
\right]
\\
g^{z}( u ) & = &
|\varepsilon|
{\rm Re}\, H_0( u )
\label{eq:plasmon-contribution-z}
\\
C_{\rm pl}
& = &
\frac{ 3\pi }{ 2 } \frac{ |\varepsilon|^2 }{ |\varepsilon + 1|^{7/2} }.
\end{eqnarray}
where $\kappa_{\rm pl} = {\rm Im}\, \gamma( K_{\rm pl} )$ is the
perpendicular
plasmon decay constant and
$H_{0,\,2}( \cdot ) \equiv H_{0,\,2}^{(1)}( \cdot )$ are Hankel
functions
(or Bessel functions of the third kind) \cite{Abramowitz}.
The superscripts $\Vert, \perp$ indicate the directions parallel and
perpendicular to the separation vector ${\bf S}$ in the interface
plane.

Eq.(\ref{eq:plasmon-contribution}) shows that the plasmon resonance
gives a contribution to the energy density that increases
exponentially
when approaching the source. This behaviour is reproduced by the
numerical evaluation of~(\ref{eq:formula-w-ij}), as shown in
fig.~2(b). As a function of the lateral distance $s$,
the correlation tensor eq.(\ref{eq:plasmon-contribution}) shows
damped oscillations whose wavelength is fixed by the plasmon wave
vector
$K_{\rm pl}$, as shown in fig.3(b). These oscillations can be made
explicit
using the asymptotic form of the Hankel function \cite{Abramowitz}
\begin{eqnarray}
&&|K_{\rm pl} s| \gg 1:
\nonumber
\\
&&
H_n( K_{\rm pl} s ) \approx
\sqrt{ \frac{ 2 }{ \pi K_{\rm pl} s } }
{\rm e}^{{\rm i}( K_{\rm pl} s - \pi / 4 - n \pi / 2 )}
\label{eq:H0-asymptotics}
\end{eqnarray}
We thus conclude that the propagation distance of the plasmon
resonance,
as contained in the imaginary part of $K_{\rm pl}$, determines
the coherence length of the field in this regime. For a dielectric
constant with small imaginary part, the inverse propagation distance
is
approximately
\begin{equation}
{\rm Im}\, K_{\rm pl}
\approx \frac{ ( {\rm Re}\, K_{\rm pl}) ^3 }{ k^2 }
\frac{ {\rm Im}\, \varepsilon }{ 2 ({\rm Re}\, \varepsilon)^2 }
\ll k
.
\label{eq:plasmon-damping}
\end{equation}
Thermally excited plasmons thus lead to a spatially coherent field
on a length scale well exceeding the vacuum wave length. They also
create a net field polarization, as shown by the anisotropy of the
tensor
elements
in~(\ref{eq:plasmon-contribution-x}-\ref{eq:plasmon-contribution-z})
[see also fig.3(b)]. This anisotropy may be understood
from the fact that the coherence between points separated by
${\bf S}$ is created by plasmons propagating parallel to this
direction, and the latter are polarized in the plane spanned
by ${\bf S}$ and the normal vector ${\bf e}_z$.

\subsection{Extreme near field: quasi-static regime}
\label{s:xnf}

We now turn to the near field limit $z \ll \lambda$. Inspecting the
integrand of~(\ref{eq:formula-w-ij}), one finds that in addition to
the
plasmon resonance, large wave vectors $K \gg k$ dominate the
integral.
This is because the exponential cutoff
${\rm e}^{- 2 z {\rm Im}\,\gamma(K)} \approx {\rm e}^{- 2 z K}$
gets effective only for $K \ge 1/z \gg k$. We thus obtain the
asymptotic
behaviour of the integral when we expand the integrand to leading
order in the limit $1/z \ge K \gg k$. The transmission coefficients,
\emph{e.g.}, become in this limit
\begin{eqnarray}
|t_{\rm p}|^2 & \approx &
\frac{ 4 |\gamma_2|^2 |\varepsilon| }{
K^2 |\varepsilon + 1|^2 }
\left(
1 + \frac{ k^2 }{ K^2 } {\rm Re}\, \frac{ \varepsilon }{ \varepsilon
+ 1 }
\right)
\nonumber\\
|t_{\rm s}|^2 & \approx &
\frac{ 4 |\gamma_2|^2  }{
K^2 }
\left(
1 + \frac{ k^2 }{ 4 K^2 } {\rm Re}\,( \varepsilon + 1 )
\right)
\label{eq:t-sp-xnf-regime}
\end{eqnarray}
We perform the integration over $K$ using eq.(\ref{eqa:J0-integral}),
as explained in appendix~\ref{a:Coulomb-trick} and
get the following asymptotic form for the diagonal elements of the
cross-correlation tensor
\begin{eqnarray}
w^{ii}( {\bf S}, z )
& \approx &
\frac{ C_{\rm xnf} }{ (k z)^3 }
g^i( s / z )
\label{eq:xnf-contribution}
\\
g^{\Vert}( u ) & = &
\frac{ 1 - u^2 / 2 }{ (1 + u^2/4)^{5/2} }
\label{eq:xnf-contribution-x}
\\
g^{\perp}( u ) & = &
\frac{ 1 }{ (1 + u^2/4)^{3/2} }
\\
g^{z}( u ) & = &
\frac{ 2 - u^2 / 4 }{ (1 + u^2/4)^{5/2} }
\label{eq:xnf-contribution-z}
\\
C_{\rm xnf} & = & \frac{ 3 }{ 32 }
{\rm Im}\, \frac{ \varepsilon - 1 }{ \varepsilon + 1 }
= \frac{ 3 }{ 16 }
\frac{ {\rm Im}\, \varepsilon  }{ |\varepsilon + 1|^2 }
\label{eq:im-stat-reflection}
\end{eqnarray}
The coherence tensor given by~(\ref{eq:xnf-contribution}) shows a
power law increase $1/z^3$ when the interface is approached, as
plotted in fig.~2(a). It therefore takes
over compared to the plasmon contribution in the `extreme near field
limit' $z \ll 1/\kappa_{\rm pl}$. In this regime, the lateral
coherence
of the field is characterized, as shown by the scale functions
$g^i(s/z)$
in~(\ref{eq:xnf-contribution-x}-\ref{eq:xnf-contribution-z}),
by a lorentzian shape
whose scale is set by the distance $z$ to the source. Hence, the
closer
one detects the field, the more it is spatially incoherent.

This behaviour may be understood from electrostatics: in the near
field,
the electromagnetic fields behave as if they were quasi-static
because they vary on a length scale much smaller than the wave length
(retardation is negligible). A near field detector is thus sensitive
to a source area of the order of $\pi z^2$, and spatial coherence
is observed when these areas overlap, hence for a separation smaller
than the distance $z$.
Similar arguments have also been put forward to interpret
subwavelength resolution in optical near field microscopy
\cite{Greffet97c,Henkel98b}. The electrostatic analogy may be pushed
even
further: it is easily checked that we get the same result
as~(\ref{eq:xnf-contribution}) using electrostatic image theory.
As a consequence of the fluctuation-dissipation theorem
\cite{Lifshitz56,Polder71,Agarwal75a,Scheel98a}, we have indeed
\begin{eqnarray}
&&
W^{ij}( {\bf r}_2, {\bf r}_1 )
\propto
{\rm Im}\, G^{ij}( {\bf r}_2, {\bf r}_1 )
\nonumber\\
&&
\propto
{\rm Im}\, E^{i}_{\rm image}( {\bf r}_2 ; \bar{\bf r}_1, \bar{\bf
d}_j )
,
\label{eq:fluct-diss}
\end{eqnarray}
where $G^{ij}$ is again the electric Green function.
The electric field ${\bf E}_{\rm image}$ is created by the image
$\bar{\bf d}_j$ of a dipole ${\bf d}_j$ (polarized along the
$x^j$-axis)
at position ${\bf r}_1$, the image dipole being located at the mirror
position $\bar{\bf r}_1 = (x_1, y_1, -z_1)$.
This image dipole field dominates the Green function
$G^{ij}$ at sufficiently close distance from the source if
the electrostatic reflection coefficient $(\varepsilon - 1 )/(
\varepsilon
+ 1 )$ has a nonzero imaginary part [see
eq.(\ref{eq:im-stat-reflection})].

We stress that there is no lower limit to the spatial coherence
length of the near field, provided one uses the framework of a local
dielectric susceptibility. Model calculations for a free
electron gas confined to a half-space show that this
framework breaks down at wave vectors $K$ of the order of the Fermi
wave vector \cite{Feibelman82,Ford84}. For our problem, this
corresponds
to typical distances of the order of $0.1\,$nm that are difficult
to achieve for near field probes even in the optical range.

\subsection{Relevance of the skin depth}
\label{s:skin}

It has become clear from the two preceding subsections that the lateral
coherence of near field radiation strongly depends on the distance of
observation to the source. It might have been expected that the skin
depth
shows up in this discussion, since it governs the penetration depth
for electric fields into the metal.
We conclude our analytical work by identifying the relevance
of this length scale.

Recall that the skin depth is given by
\begin{equation}
\delta = \frac{ \lambda / 2\pi }{ {\rm Im}\, \sqrt{\varepsilon} }
\label{eq:def-skin-depth}
\end{equation}
For good conductors and low frequencies, the dielectric function
is dominated by its zero-frequency pole
\begin{equation}
\varepsilon = \frac{ {\rm i} \sigma }{ \varepsilon_0 \omega }
\label{eq:def-sigma}
\end{equation}
where $\sigma$ is the (possibly frequency dependent) conductivity.
In particular, one has $|\varepsilon| \gg 1$ and $\delta \ll \lambda$
in this regime:
the source material hence approaches a perfect conductor.
This implies that the
large $K$ expansion of the transmission coefficients
in~(\ref{eq:formula-w-ij})
has to be reconsidered: while the limit $K \gg k$ may be justified,
the limit $K \gg k |\sqrt{\varepsilon}|$ may be not. We find that
there exists an intermediate distance regime $\delta \ll z \ll
\lambda$
(corresponding to wave vectors $k|\sqrt{\varepsilon}| \gg K \gg k$)
where the coherence tensor shows
a different behaviour \cite{Henkel99c}. The expansion
of the transmission coefficients in this regime reads
\begin{eqnarray}
|t_{\rm p}|^2 & \approx &
\frac{ 4 |\gamma_2|^2 }
{ |\varepsilon| K^2 }
\left[
1 + {\mathcal O}\left( \frac{ K }{ k \sqrt{\varepsilon} } \right)
\right]
\nonumber\\
|t_{\rm s}|^2 & \approx &
\frac{ 4 |\gamma_2|^2  }
{ |\varepsilon| k^2 }
\left[
1 + {\mathcal O}\left( \frac{ K }{ k \sqrt{\varepsilon} } \right)
\right]
\label{eq:t-sp-skin-regime}
\end{eqnarray}
We finally get an isotropic coherence tensor
\begin{eqnarray}
w^{ii}( {\bf S}, z )
& \approx &
\frac{ 3 }{ 16 }
\frac{ \delta }{ k z^2 }
g( s / z )
\label{eq:skin-contribution}
\\
g( u ) & = &
\frac{ 1 }{ (1 + u^2/4)^{3/2} }
\end{eqnarray}
The skin layer dominated regime is thus characterized by a $1/z^2$
power
law for the energy density.  As shown in figure~4(a), the skin depth
$\delta$
separates this regime from the extreme near field where a different
power law $1/z^3$ prevails.
\begin{figure*}
\centerline{
\resizebox{0.4\textwidth}{!}{
\includegraphics*{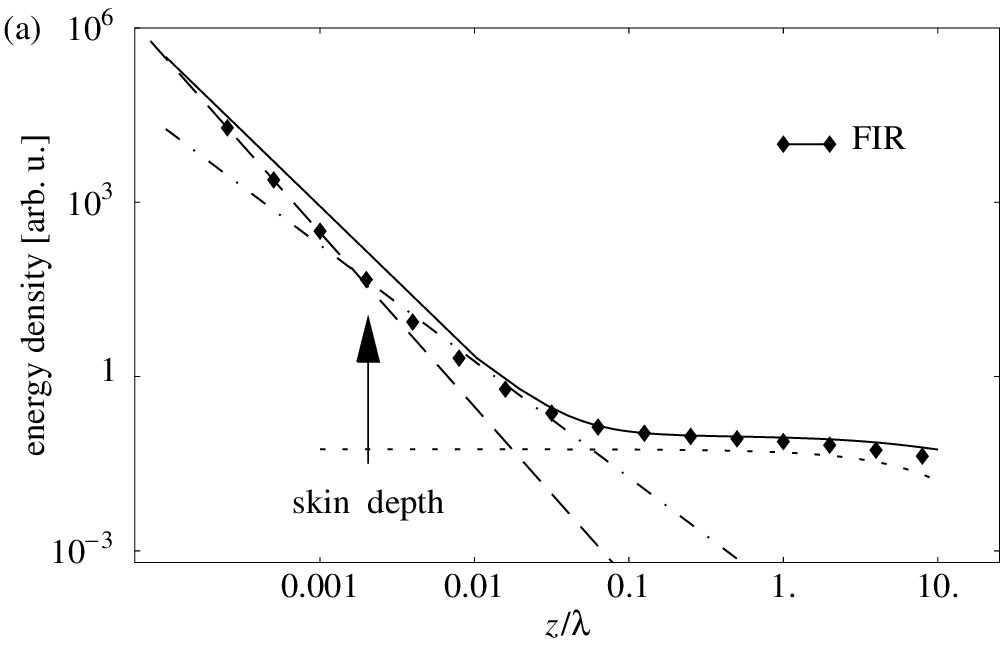}}
\hspace*{3mm}
\resizebox{0.4\textwidth}{!}{
\includegraphics*{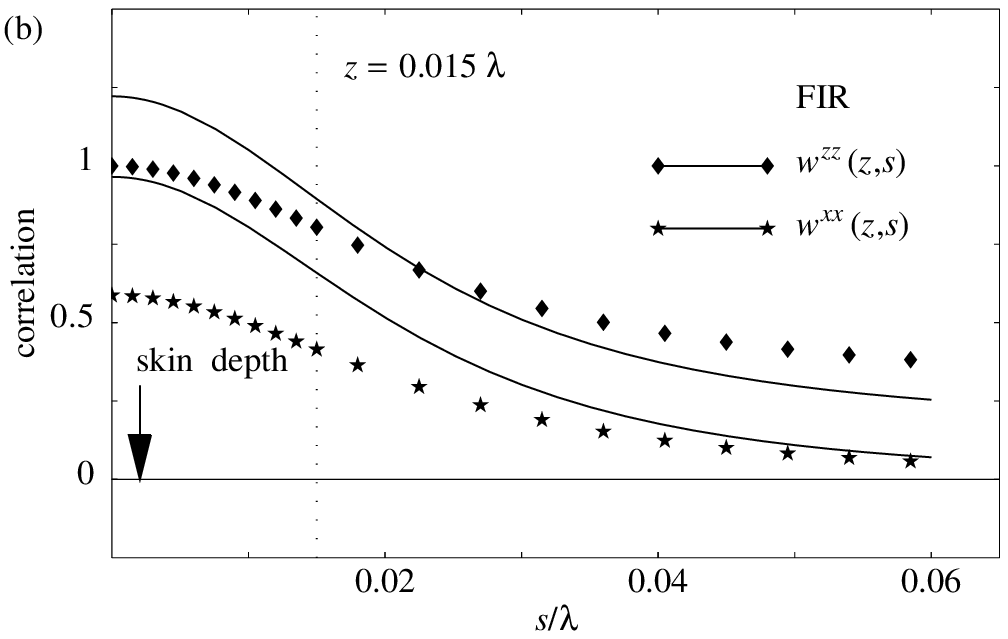}}
}
\caption[]{Near field coherence in the skin-effect dominated
regime $\delta \ll z \ll \lambda$, typical for metals in the
far infrared. Dots: numerical
evaluation of the integral~(\ref{eq:def-w-ij}), lines:
analytic asymptotics discussed in the text.
\\
(a):
energy density ${\rm Tr}\,w^{ij}( {\bf S} = {\bf 0}, z )$
above a metallic surface with $\varepsilon = -8.26 + {\rm i}\,10^4$
(gold extrapolated to $\lambda = 3.7\,\mu$m). The solid line
is the sum of the asymptotic contributions derived in this paper.
(b):
normalized lateral coherence functions
$w^{ii}( {\bf S}, z ) / w^{zz}( {\bf 0}, z )$
($i = x,\,z$)
for fixed distance $z$, plotted vs.\ the lateral separation
$s = |{\bf S}|$. The $x$- and $z$-polarizations differ because
the plasmon contribution already comes into play. The numerically
computed $w^{zz}( {\bf 0}, z )$ was used to normalize
all curves, this is why the analytic correlations exceed unity.
}
\end{figure*}
We observe from fig.4(b) and~(\ref{eq:skin-contribution})
that the lateral coherence length is equal to the distance $z$
from the source, as in the extreme near field regime. This is not
so surprising since the field propagation in the vacuum half
space above the source is governed by the length scales $\lambda$
and $z$, whatever the smaller, while the skin depth is only relevant
for the propagation inside the source.

To conclude, we recall that the different contributions to the
correlation tensor originate in distinct domains on the $K$-axis in
the integral~(\ref{eq:def-w-ij}). The total correlation tensor
is therefore given by the sum of the surface plasmon, extreme near
field, and skin-layer contributions. The accuracy of this approximation
is visible in figs.2,~4(a). Note that in the figures, the numerically 
computed far-field energy density has been added to get the correct 
large distance limit.

\section{Conclusion}

In the near field, the spatial coherence of thermal radiation
differs strongly from the blackbody field. Confined field modes like
surface plasmon polaritons that
propagate along the source surface make the field spatially
coherent over large scales if they dominate the radiation density.
At close distance (smaller than the skin depth), the radiation
is dominated by quasi-static fields, and the coherence length drops
well below the wave length, being limited only by the (non)locality 
of
the dielectric response of the source material. The cross over
between these
regimes is determined by the skin depth and the electrostatic
reflection
coefficient. We conclude that in the near field,
macroscopic concepts like a local emissivity are still meaningful
at the subwavelength scale, provided coherent surface excitations
are absent or subdominant.

The asymptotic forms for the cross spectral density tensor obtained
in this paper are useful to characterize thermal noise fields that
may perturb particles in integrated microtraps `mounted' with
electromagnetic fields above a solid surface
\cite{Dowling97,Balykin99,Schmiedmayer98a,%
Hinds00,Reichel99,Schmiedmayer00a,Prentiss00,Anderson99}. The
concomitant
scattering and decoherence of the guided matter waves is discussed
elsewhere \cite{Henkel00a,Henkel00c}.

The fluctuation electrodynamics used in this paper enabled us to
treat
a non-equilibrium situation (thermal source in vacuum at $T=0$) where
the fluctuation-dissipation theorem for the electric field is not
immediately applicable. In particular, we neglected the zero-point
radiation impinging on the interface from the empty half-space. The
domain of validity of this approximation, as well as the calculation
of anti-normal ordered correlation functions will be the subject of
future investigations.

\paragraph{Acknowledgements.}
C.H. thanks S. Scheel for many fruitful exchanges.

\appendix

\section{Notations for the plane interface}
\label{a:Green}

The Green tensor describing the emission from the source $z'<0$ into
the vacuum
half space $z>0$ may be written in spatial Fourier space as
\cite{Agarwal75a,Maradudin75}
\begin{eqnarray}
&&
G^{ij}( {\bf r}, {\bf r}' )
=
\int\!\frac{ {\rm d}^2{\bf K} }{ (2\pi)^2 }
{\rm e}^{ {\rm i}{\bf K}\cdot( {\bf R} - {\bf R}' ) }
G^{ij}( {\bf K}, z, z' )
\nonumber
\\
&&
G^{ij}( {\bf K}, z, z' )
=
\frac{ {\rm i} k^2 }{ 2\varepsilon_0 \gamma_2 }
\nonumber
\\
&&
\times\sum_{\mu = {\rm s},\, {\rm p}}
{e}^{(t)}_{\mu,i} {e}^{(2)}_{\mu,j} t_\mu
{\rm e}^{{\rm i} ( \gamma z - \gamma_2 z' ) }
\label{eqa:Green-tensor}
\end{eqnarray}
We use bold capitals to denote vectors parallel to the interface,
\emph{e.g.},
${\bf K} = (k_x,\, k_y,\, 0)$.
The vertical components of the wave vectors in vacuum and
inside the source are, respectively,
\begin{eqnarray}
\gamma & = & +\sqrt{ k^2 - {\bf K}^2 }, \qquad
{\rm Im}\,\gamma > 0
\\
\gamma_2 & = & +\sqrt{ \varepsilon k^2 - {\bf K}^2 }, \qquad
{\rm Im}\,\gamma_2 > 0
\end{eqnarray}
The polarization vectors for the s- (or TE-) and p- (TM-)polarized
waves are
taken as
\begin{eqnarray}
{\bf e}^{(t)}_{\rm s} & = &
{\bf e}^{(2)}_{\rm s} =
\hat{\bf K} \times \hat{\bf e}_z
\\
{\bf e}^{(t)}_{\rm p} & = &
\frac{ K \hat{\bf z} - \gamma \hat{\bf K} }{  k }
\\
{\bf e}^{(2)}_{\rm p} & = &
\frac{ K \hat{\bf z} - \gamma_2 \hat{\bf K} }{ \sqrt{\varepsilon} k }
\label{eqa:pol-vectors}
\end{eqnarray}
where $\hat{\bf K}$ is the unit vector parallel to ${\bf K}$.
Finally, with this choice for the polarization vectors, the Fresnel
transmission coefficients are
\begin{eqnarray}
t_{\rm s} & = &
\frac{ 2 \gamma_2 }{
\gamma + \gamma_2 },
\qquad
t_{\rm p} =
\frac{ 2 \gamma_2 \sqrt{ \varepsilon } }{
\varepsilon \gamma + \gamma_2 }
\label{eqa:def-t-sp}
\end{eqnarray}

When the Green tensor~(\ref{eqa:Green-tensor}) is inserted into the
integral~(\ref{eq:SE-and-Green}), the spatial integration over ${\bf
R}'$
yields a $\delta$-function for the lateral wave vectors.
The integration over $z'$ is then
\begin{equation}
\int\limits_{-\infty}^{0}\!{\rm d}z' \,
{\rm e}^{-{\rm i}( \gamma_2 - \gamma^*_2 ) z' }
=
\frac{ 1 }{ 2 \, {\rm Im}\,\gamma_2 }
\label{eq:result-z-integral}
\end{equation}
where the convergence is ensured by the positive imaginary part of
$\gamma_2$. The resulting coherence tensor is then of the
form~(\ref{eq:def-w-ij}). We use the identity
\begin{equation}
k^2 {\rm Im}\,\varepsilon
= 2 \, {\rm Im}\, \gamma_2 \, {\rm Re}\, \gamma_2
\end{equation}
and get after some elementary algebra:
\begin{eqnarray}
w^{ij}( {\bf K} ) & = &
\frac{ 3\pi }{ 4 k }
\frac{ {\rm Re}\, \gamma_2 }{ |\gamma_2|^2 }
\times{}
\nonumber\\
&& {} \times
\sum_\mu
e^{(t)}_{\mu,i}\,
e^{(t)*}_{\mu,j}\,
|{\bf e}^{(2)}_{\mu}|^2
|t_{\mu}|^2
\label{eq:formula-w-ij}
\end{eqnarray}

\section{Components of the coherence tensor}

In this appendix, we outline the calculation for the components
of the coherence tensor.

\subsection{Angular integrations}
\label{a:plasmon-polarization}

The only quantities in~(\ref{eq:formula-w-ij})
that depend on the
angle $\varphi$ between the lateral wavevector ${\bf K}$ and the
separation ${\bf S}$ are the polarization vectors ${\bf e}_\mu$.
To simplify the calculation, we choose the $x$-axis parallel to
${\bf S}$. We then get the following azimuthal integrals
(eq.9.1.18 of \cite{Abramowitz})
\begin{equation}
\int_{-\pi}^{\pi}\! \frac{ {\rm d}\varphi }{ \pi } \,
{\rm e}^{ {\rm i} K s \cos\varphi }
\left(
\begin{array}{c}
\sin^2\varphi\\
\cos^2\varphi
\end{array}
\right)
=
J_0( K s )
\pm
J_2( K s )
\label{eq:angular-trick}
\end{equation}
The integrals with $\sin\varphi\,\cos\varphi$ vanish due to parity.
We also note that one also gets nonzero off-diagonal elements
$W^{xz}, \, W^{zx}$ due to p-polarized modes. For simplicity, these
are not discussed here.

\subsection{Radial integrations}

We are left with integrals over the radial wave vector $K$.
These are worked out using the
definitions~(\ref{eqa:pol-vectors}) of the polarization vectors
and the transmission coefficients~(\ref{eqa:def-t-sp}).

\subsubsection{Plasmon pole}
\label{a:contour}

To find the plasmon contribution, we
extract, as mentioned in the main text, the pole of the $t_{\rm p}$
coefficient and approximate the other factors by their values at the
pole. The remaining integral can be reduced to the following standard
form
\begin{equation}
\int_0^\infty \!\frac{ x\,{\rm d}x }{ x^2 - q^2 } J_0( x s )
= \frac{ {\rm i} \pi }{ 2 }
H_0^{(1)}( q s )
\label{eq:J0-theorem}
\end{equation}
for ${\rm Im}\, q > 0, \: s > 0$. To prove this identity, we use
contour integration. The Bessel function is written as
(\cite{Abramowitz}, eqs.~9.1.3, 9.1.39)
\begin{equation}
J_0( x ) = \frac12 \left[ H_0^{(1)}( x ) -
H_0^{(1)}( {\rm e}^{{\rm i}\pi } x ) \right]
\label{eqa:contour1}
\end{equation}
where $H_0^{(1)}( x )$ is the Hankel function.
The integral may now be written as
\begin{equation}
\frac12 \int_{{\rm e}^{{\rm i}\pi} \infty}^\infty
\!\frac{ x\,{\rm d}x }{ x^2 - q^2 } H_0^{(1)}( x s )
\label{eqa:contour2}
\end{equation}
with an integration path running just above the negative real axis.
The
Hankel function is analytic in the upper half plane and vanishes
exponentially for $|x|\to\infty$ there [see
eq.(\ref{eq:H0-asymptotics})].
Therefore, closing the integration contour with a half circle,
the integral is given by the residue at the pole
$x = +q$ (because ${\rm Im}\,q > 0$), and we get
\begin{equation}
\int_{{\rm e}^{{\rm i}\pi} \infty}^\infty
\!\frac{ x\,{\rm d}x }{ x^2 - q^2 } H_0^{(1)}( x s )
=
{\rm i}\pi H_0^{(1)}( q s )
\end{equation}
This proves~(\ref{eq:J0-theorem}). Taking the imaginary part, we find
both the trace and the
$zz$-component~(\ref{eq:plasmon-contribution-z})
of the correlation tensor~(\ref{eq:plasmon-contribution}).

For the $xx$- and $yy$ components of the coherence tensor, we also
need
the integral~(\ref{eq:J0-theorem}) with the Bessel function $J_2( x s
)$
instead of $J_0( x s )$ [cf.\ eq.(\ref{eq:angular-trick})].
Using the same reasoning as above, this integral is transformed into
\begin{equation}
\frac12 \int_{{\rm e}^{{\rm i}\pi} \infty}^\infty
\!\frac{ x\,{\rm d}x }{ x^2 - q^2 } H_2^{(1)}( x s )
\label{eqa:contour4}
\end{equation}
In addition to the pole at $x = q$, we now have a contribution from
the
$-4{\rm i}/(xs)^2$ singularity of the Bessel $H_2^{(1)}( x s )$
function at
the origin. This singularity lies on the integration path and is
therefore
taken into account by half its (negative) residue at $x = 0$.
Combining the latter with the residue at $x = q$, we get
\begin{equation}
\int_{{\rm e}^{{\rm i}\pi} \infty}^\infty
\!\frac{ x\,{\rm d}x }{ x^2 - q^2 } H_2^{(1)}( x s )
= {\rm i}\pi
H_2^{(1)}( q s ) - \frac{ 4 \pi }{ q^2 s^2 }
\label{eqa:contour5}
\end{equation}
We may verify the sign of the second term by checking that the
function~(\ref{eqa:contour5}) vanishes in the limit $s \to 0$,
as is the case for the left hand side of~(\ref{eqa:contour4}).

\subsubsection{Near field regime}
\label{a:Coulomb-trick}

In the near field regimes $K \gg k|\sqrt{\varepsilon}|$ (extreme near
field) and $k|\sqrt{\varepsilon}| \gg K \gg k$ (skin layer dominated
regime), the expansions~(\ref{eq:t-sp-xnf-regime},
\ref{eq:t-sp-skin-regime})
of the transmission coefficients are straightforward to obtain.
The final integration involves integer powers
of $K$ times products of Bessel functions and exponentials and is
performed using the following identity
\begin{equation}
\int_0^{\infty}\!{\rm d}K \, J_0( K s ) {\rm e}^{ - 2 K z }
=
\frac{ 1 }{ \sqrt{ 4 z^2 + s^2 } }
\label{eqa:J0-integral}
\end{equation}
This may be proven starting from the Fourier expansion of the Coulomb
potential (writing ${\bf k} = ( {\bf K}, k^z )$)
\begin{equation}
\frac{ 1 }{ r } = \frac{ 1 }{ 2\pi^2 }
\int\!
{\rm d}^2 {\bf K} \, {\rm d}k^z
\,\frac{
{\rm e}^{ {\rm i} {\bf k} \cdot {\bf r} }
}{ {\bf k}^2 }
\label{eqa:Coulomb-Fourier}
\end{equation}
and evaluating the integral over the vertical wave vector component
$k^z$
with contour integration (for $z>0$, a single pole at
$k^z = {\rm i}|{\bf K}|$ contributes). The derivatives
of~(\ref{eqa:J0-integral}) with respect to $z$ and $s$ then provide
all
necessary integrals.

\bibliography{/Net/Users/carstenh/Biblio/Database/journals,/Net/Users/carstenh/Biblio/Database/biblioac,/Net/Users/carstenh/Biblio/Database/bibliodh,/Net/Users/carstenh/Biblio/Database/biblioio,/Net/Users/carstenh/Biblio/Database/bibliopz}

\begin{thebibliography}{10}

\bibitem{MandelWolf}
L. Mandel and E. Wolf, Optical coherence and quantum optics (Cambridge
  University Press, Cambridge, 1995).

\bibitem{Greffet99}
R. Carminati and J.-J. Greffet, Phys. Rev. Lett. 82 (1999) 1660.

\bibitem{Shchegrov00}
A.~V. Shchegrov, K. Joulain, R. Carminati, and J.-J. Greffet, Phys. Rev. Lett.
  (2000), in press.

\bibitem{Lifshitz56}
E.~M. Lifshitz, Soviet Phys. JETP 2 (1956) 73, [J. Exper. Theoret. Phys.
  USSR {\bf 29}, 94 (1955)].

\bibitem{Polder71}
D. Polder and M.~V. Hove, Phys. Rev. B 4 (1971) 3303.

\bibitem{Schwinger78}
J. Schwinger, J. Lester L.~DeRaad, and K.~A. Milton, Ann. Phys. (N.Y.) 115
  (1978) 1.

\bibitem{Rytov3}
S.~M. Rytov, Y.~A. Kravtsov, and V.~I. Tatarskii, Principles of Statistical
  Radiophysics (Springer, Berlin, 1989), Vol.~3, Chap.~3.

\bibitem{Feibelman82}
P.~J. Feibelman, Progr. Surf. Sci. 12 (1982) 287.

\bibitem{Ford84}
G.~W. Ford and W.~H. Weber, Phys. Rep. 113 (1984) 195.

\bibitem{Agarwal75a}
G.~S. Agarwal, Phys. Rev. A 11 (1975) 230.

\bibitem{Scheel98a}
S. Scheel, L. Kn\"{o}ll, and D.-G. Welsch, Phys. Rev. A 58 (1998) 700.

\bibitem{Maradudin75}
A.~A. Maradudin and D.~L. Mills, Phys. Rev. B 11 (1975) 1392.

\bibitem{Henkel98a}
C. Henkel and J.-Y. Courtois, Eur. Phys. J. D 3 (1998) 129.

\bibitem{Palik}
E. Palik, Handbook of optical constants of solids (Academic, San Diego, 1985).

\bibitem{AgranovichMills}
Surface Polaritons, edited by V.~M. Agranovich and D.~L. Mills (North-Holland,
  Amsterdam, 1982).

\bibitem{Raether}
H. Raether, Surface Plasmons on Smooth and Rough Surfaces and on Gratings,
  Vol.~111 of Springer Tracts in Modern Physics (Springer, Berlin, Heidelberg,
  1988).

\bibitem{Abramowitz}
Handbook of Mathematical Functions, ninth ed., edited by M. Abramowitz and
  I.~A. Stegun (Dover Publications, Inc., New York, 1972).

\bibitem{Greffet97c}
J.-J. Greffet and R. Carminati, Progr. Surf. Sci. 56 (1997) 133.

\bibitem{Henkel98b}
C. Henkel and V. Sandoghdar, Opt. Commun. 158 (1998) 250.

\bibitem{Henkel99c}
C. Henkel, S. P{\"o}tting, and M. Wilkens, Appl. Phys.~B 69 (1999) 379.

\bibitem{Dowling97}
J.~P. Dowling and J. Gea-Banacloche,  in Adv. At. Mol. Opt. Phys., edited by
  P.~R. Berman (Academic Press, New York, 1997), Vol.~37, {\rm Suppl.~3}, pp.\
  1--94.

\bibitem{Balykin99}
V.~I. Balykin,  in Adv. At. Mol. Opt. Phys., edited by B. Bederson and H.
  Walther (Academic, San Diego, 1999), Vol.~41, pp.\ 181--260.

\bibitem{Schmiedmayer98a}
J. Schmiedmayer, Eur. Phys. J. D 4 (1998) 57.

\bibitem{Hinds00}
M. Key {\it et~al.}, Phys. Rev. Lett. 84 (2000) 1371.

\bibitem{Reichel99}
J. Reichel, W. H\"ansel, and T.~W. H\"ansch, Phys. Rev. Lett. 83 (1999) 3398.

\bibitem{Schmiedmayer00a}
R. Folman, P. Kr\"uger, D. Cassettari, B. Hessmo, T. Maier, and J.
  Schmiedmayer, Phys. Rev. Lett. 84 (2000) 4749.

\bibitem{Prentiss00}
N.~H. Dekker {\it et~al.}, Phys. Rev. Lett. 84 (2000) 1124.

\bibitem{Anderson99}
D. M\"uller, D.~Z. Anderson, R.~J. Grow, P.~D.~D. Schwindt, and E.~A. Cornell,
  Phys. Rev. Lett. 83 (1999) 5194.

\bibitem{Henkel00a}
C. Henkel and M. Wilkens, acta phys. slov. 50 (2000) 293
[quant-ph/0005038].

\bibitem{Henkel00c}
C. Henkel and S. P\"otting, ``Coherent transport of matter waves'',
submitted for publication in Appl. Phys. B. [quant-ph/0007083]

\end{thebibliography}
\bibliographystyle{/Net/Users/carstenh/Biblio/Database/bst/optcomm}

\end{multicols}

\end{document}